# Probing Electrical Properties of A Silicon Nanocrystal Thin Film Using X-ray Photoelectron Spectroscopy

Amrit Laudari[1], Sameera Pathiranage[1], Salim A. Thomas[2], Reed J. Petersen[3], Kenneth J. Anderson[2], Todd A. Pringle[2], Erik K. Hobbie[2], Nuri Oncel[1]

[1] Department of Physics and Astrophysics, University of North Dakota, 58202, Grand Forks, North Dakota, USA

[2] Materials and Nanotechnology Program, North Dakota State University, 58108, Fargo, North Dakota, USA

[3] Department of Physics, North Dakota State University, 58108, Fargo, North Dakota, USA

The author to whom correspondence may be addressed: nuri.oncel@und.edu

## Abstract

We performed X-ray photoelectron spectroscopy (XPS) measurements on a thin film of Si nanocrystals (SiNCs) while applying DC or AC external biases to extract the resistance and the capacitance of the thin film. The measurement consists of the application of $10\ V$ DC or square wave pulses of $10\ V$ amplitude to the sample at various frequencies ranging from $0.01\ Hz$ to $1\ MHz$ while recording X-ray photoemission data. To analyze the data, we propose three different models with varying degrees of accuracy. The calculated capacitance of SiNCs agrees with the experimental value in the literature.

## Introduction

X-ray photoelectron spectroscopy (XPS) is a powerful tool for extracting the chemical, physical, and electrical properties of surfaces and interfaces.[1–7] Since XPS is a noncontact and mostly a non-destructive technique, it has been used to probe the electrical properties of materials while maintaining the inherent chemical specificity of XPS. Ideally, during an XPS measurement, the net sum of electrons coming into and leaving the sample should be zero, leaving the sample charge neutral. This is achieved by grounding the sample to the spectrometer ground. Therefore, any changes in binding energies are attributed to the chemical potential and polarizability of compounds. These chemical shifts are used to identify the state of the elements. When a DC or an AC external bias is applied to a conducting sample, the shifts in binding energies are equal to the applied external bias. However, for insulating samples, the grounding the sample cannot prevent the accumulation of positive charges on the surface which causes differential charging. For a long time, differential charging has been considered a hindrance. Electron flood guns have been utilized to control and reduce the effect, but the complete elimination of differential charging is difficult, and if not done properly it can cause the accumulation of negative charges on the surface leading to erroneous interpretation. However, differential charging can also be employed to study the structural and electronic properties of dielectric films, where equivalent circuit models have been proposed to extract the capacitance and resistance from the observed shifts in binding energy with applied external AC or DC voltage.[8,9] These previous attempts to determine capacitance rely on time-dependent protocols with external control of the applied bias and analyzer by using data acquisition cards.[5] Typically, four steps are repeated at every energy point: a positive bias is applied, the system stabilizes, the voltage is abruptly reversed, and electrons are counted. If the time constant ($\tau$) of the equivalent circuit is comparable to the observation (dwell) time, we anticipate an exponential decay, $e^{-t/\tau}$. The experimentally limiting factor is the dwell time, around 5 ms for XPS. To have reasonable signal/noise ratio limits, the dwell time cannot be randomly small. The sufficient observation time also strongly depends on the material being analyzed and the quality of the analyzer, and in many cases, it is not possible to improve. Therefore, this technique has limitations when the time constant of the RC circuit is comparable to the dwell time. To overcome these issues, an alternative method was developed to extract the capacitance of the thin film. The advantage of this method is that closed-form mathematical equations were derived to model binding energy shifts as a function of applied bias. By using this method, a basic curve fitting algorithm can be used to extract the resistances and capacitances of thin films.

Si nanocrystals (SiNCs) were chosen to test the proposed analysis method. Semiconductor nanocrystals have been at the epicenter of recent research for a wide range of potential applications ranging from photovoltaics and fluorescent contrast agents to color displays such as QLED TVs.[10–12] Among the semiconductor nanocrystals, SiNCs have found possible applications in niche technologies such as bio-labeling and light-harvesting because of their nontoxic nature. The optical properties of these nanocrystals strongly depend on their coupling to the substrate. Studies show that electrical properties of SiNCs (e.g., the dielectric function) show a significant suppression, as compared to bulk crystalline silicon, due to the quantum size effect.[13] The estimation of electrical parameters using XPS measurements has been previously reported by various groups.[14–17] However, only a few studies focused on the electrical properties of thin films of SiNCs.[18] As an alternative to intricate time-dependent XPS measurements, in this paper we demonstrate a relatively straightforward method of analysis that does not require modifications to an existing XPS system.

## Experimental Methods

A thin film of Ag was deposited from a tungsten crucible on a freshly peeled mica sheet in a custom-made high-vacuum system. The quality of the polycrystalline Ag surface was confirmed by using the NanoMagnetics Instruments ezAFM system. For surface passivation, SiNCs were collected on stainless steel mesh and transferred to microwave reaction vials containing mesitylene and 1-dodecene (at a 5:1 ratio) and heated to 185 °C for 1-2 hours. Details related to the synthesis of the SiNCs are reported elsewhere.[19] The sample was washed repeatedly in methanol to remove unbound 1-dodecene. The average size of the SiNCs is 7 nm, with a broad photoluminescence peak at 940 nm and a PL quantum yield of 50 %. Colloidal SiNCs were spin-coated at the rate of about 1600 rpm for 60 seconds with dynamic dispatch on polycrystalline Ag film using a homemade spin coater. A PHI Model 10-360 electron spectrometer with non-monochromatized Al Kα (1486.3 eV) X-rays were used for XPS analysis. During the XPS measurements, the pressure in the chamber was kept at or below $1 \times 10^{-9}$ mbar. All XPS core-level spectra were analyzed using Augerscan software equipped with its curve-fitting program. The core-level peaks were fitted using a Gaussian–Lorentzian (GL) function to include the instrumental response function along with the core level line shape. The secondary-electron background was subtracted using a Shirley function. A thin foil of Au is attached at the corner of SiNCs sample and grounded directly via contact with the sample holder. When the sample was grounded. the binding energies of Si 2$p$ and Au 4$f$ were measured at 100.3 eV and 83.5 eV respectively. These binding energies were used as a reference. We applied both DC voltage and square wave pulses to experimentally determine the resistance and capacitance of the thin film of SiNCs. An Agilent 33500B waveform generator was used to apply a 10 V DC bias and 0 to 10 V square wave pulses. The duty cycle of the applied square waves was 50% with frequencies varying from 0.01 Hz to 1 MHz. To calculate the fitting of the experimental data, we used the NonlinearModelFit function in Wolfram Mathematica software.

## Results and Discussions

We interpret the measurements using an equivalent circuit model of the system. (See Figure 1a and 1b) To simulate a real capacitor, two resistances, one series, and one parallel to the capacitor, were used. An ideal capacitor stores and releases electrical energy without dissipation. However, all real capacitors have imperfections that create internal resistance and leakage. The series resistance represents the internal resistance of a real capacitor whereas the parallel resistance carries the leakage current. A constant flow of photoelectrons out of the sample surface and the free electrons in the vacuum are represented in the proposed model as a voltage-controlled current source, $I_p$.[20]

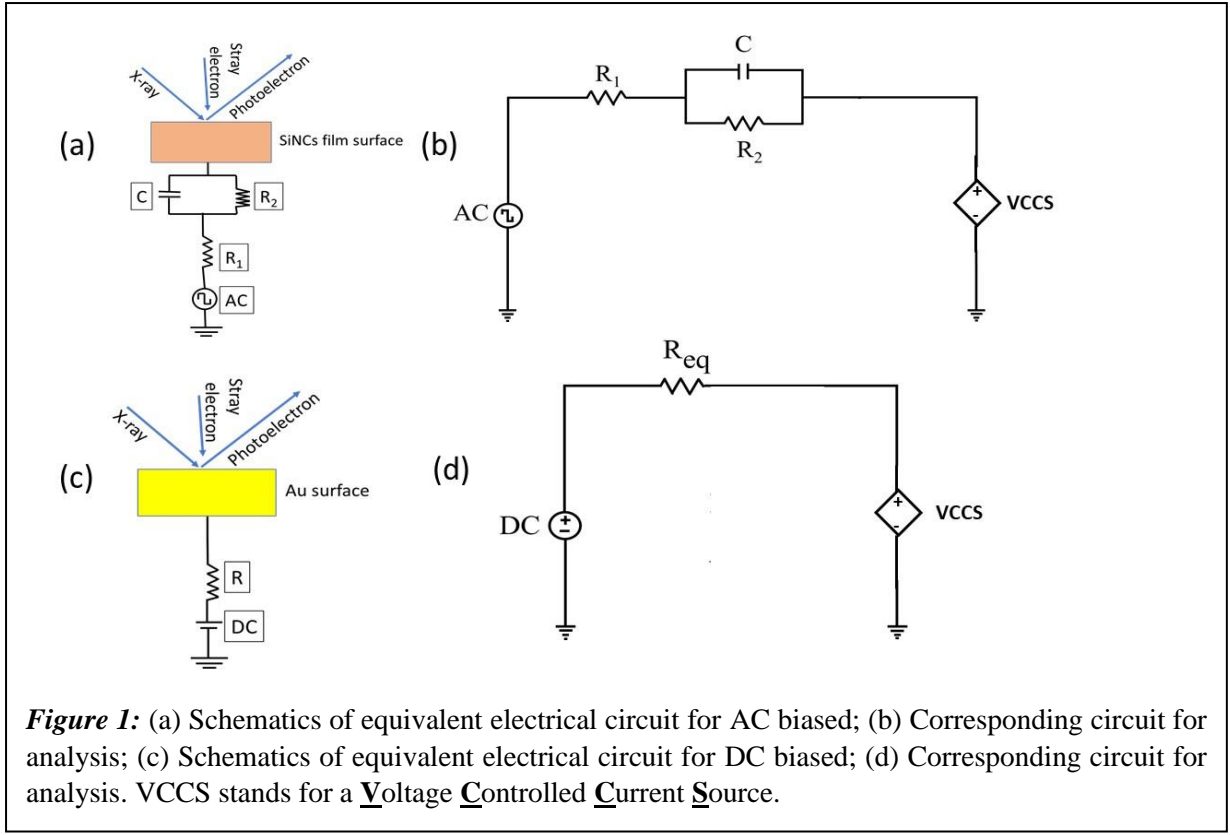

*Figure 1:* (a) Schematics of equivalent electrical circuit for AC biased; (b) Corresponding circuit for analysis; (c) Schematics of equivalent electrical circuit for DC biased; (d) Corresponding circuit for analysis. VCCS stands for a **V**oltage **C**ontrolled **C**urrent **S**ource.

When the SiNC sample was grounded, the Si 2*p* peak was measured at 100.3 eV. On the other hand, when an external $+10\,V\,DC$ voltage was applied, the binding energy of Si 2*p* peak shifted $+6.2$ eV as shown in Figure 2a. When the same experiment was repeated on the Au foil, as expected, the shift in the binding energy of Au 4*f* was $+10\,eV$ (Figure 2b). When a DC voltage is applied, the part of the circuit representing the sample can be simplified to an equivalent resistor $R_{eq} = R_1 + R_2$ since DC voltage cannot sustain a current through an ideal capacitor (Figure 1c and 1d). The equivalent resistor causes the voltage to drop across the sample and therefore lowers the shift in the binding energy. Therefore, to determine $R_{eq}$, we repeated the DC measurements on an Au foil by using an external variable resistor. A 280 MΩ external resistor shifted the Au 4*f* peak $+6.2\,eV$ when $+10\,V$ external bias was applied. Under these conditions, the current passing through the sample was ~13.6 nA.

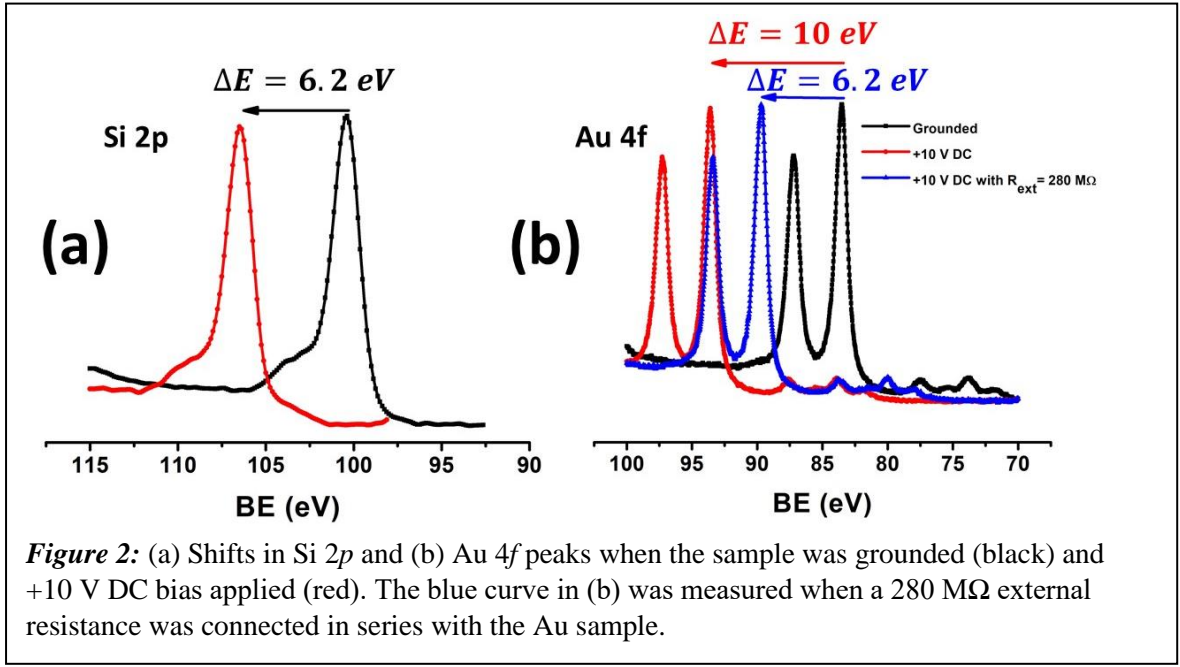

*Figure 2:* (a) Shifts in Si 2*p* and (b) Au 4*f* peaks when the sample was grounded (black) and +10 V DC bias applied (red). The blue curve in (b) was measured when a 280 MΩ external resistance was connected in series with the Au sample.

To extract the capacitance of the thin film, we performed a sequence of XPS measurements with a square-wave frequency ranging from $0.01\,Hz\,to\,10^6$ Hz. In Figure 3, the graph with solid triangles shows that the shift in the binding energy of Si 2*p* peak exhibits a sigmoidal frequency dependence, characteristic of an RC circuit with an AC power supply as described by Eq.1. We assumed that the current flowing through the sample remains constant at 13.6 nA in the dynamic range and used Eq.1 to fit the experimental data and determine $R_1, R_2,$ and $C$ (Table-1, 1st row). In Figure-3, the graph with open squares shows the fitted curve using Eq. 1. It is possible to argue that Eq.1 would not be adequate to describe the circuit since the experiments were done using square wave pulses rather than an AC source at a fixed frequency ($\omega$) varying sinusoidally with time. To improve the accuracy of the fitting, we used the Fourier series to construct square wave pulses as harmonically related sinusoids. (Eq.2) The shift in the binding energy at each frequency was calculated by a weighted summation of Fourier components. (Eq.3) By using Eq.3, we extracted $R_1, R_2,$ and $C$ values. (Table-1, 2nd row) In Figure 3, the plot with open circles shows the fitted curve. A comparison of both results shows that using the Fourier series did not improve the accuracy of the fitting significantly as fitted curves look almost identical. The extracted values for $R_1, R_2$, and $C$ using Eq.1 and Eq.3 were also within the standard error of each model. On an important note, in both cases, the models fail to reproduce experimental data at low frequencies. Another curious point to note is that the size of equivalent resistance, $R_{eq}$, extracted using Eq.1 and Eq.3 undershoot the experimentally determined $R_{eq} = 280$ MΩ.

$$\Delta V_\omega = 10\ V - I\left(R_1 + \frac{1}{\sqrt{\frac{1}{R_2^2}+(\omega C)^2}}\right) \quad \text{Eq.1}$$

$$V_\omega(t) = I\left[0.5 + \sum_{n=1}^{\infty}\left(\frac{-1}{n\pi}\right)(cos(n\pi) - 1)\ sin(n\omega t)\right] \quad \text{Eq.2}$$

$$\Delta V_\omega = 10\ V - I\left[0.5 + \sum_{n=1}^{\infty}\left(\frac{-1}{n\pi}\right)(cos(n\pi) - 1)\left(R_1 + \frac{1}{\sqrt{\frac{1}{R_2^2}+(n\omega C)^2}}\right)\right] \quad \text{Eq.3}$$

|  | $R^2$ | $R_1$ (MΩ) | $R_2$ (MΩ) | $R_{eq}$ (MΩ) | C (pF) | $V_{sq}$ (Volts) | τ (seconds) |
|---|---|---|---|---|---|---|---|
| Equation 1 | >0.99 | 179 ± 3.8 | 86 ± 7.2 | 265 ± 8.14 | 80.7 ± 27 | N.A. | N.A. |
| Equation 3 | >0.99 | 97.8 ± 7.3 | 136 ± 11.3 | 234 ± 13.5 | 51 ± 17 | N.A. | N.A. |
| Equation 6 | >0.99 | N.A. | N.A. | 280 | 41 ± 12.5 | 1.33 ± 0.048 | N.A. |
| Equation 8 | >0.99 | 172 ± 2.2 | 98 ± 3.9 | 270 ± 4.47 | 125.4 ± 21.5 (DC capacitance) | N.A. | 0.0033 ± 0.0008 |

Table-1: The resistances and capacitance of the equivalent circuit calculated by using three different methods.

To extract the capacitance of the thin film, we propose a time-averaged approach.[20] When a square wave pulse is applied to an RC circuit at a sufficiently slow frequency (with respect to the time constant of the circuit), the capacitor gets fully charged during the on-cycle and fully discharges during the off-cycle with an exponential decay $e^{-t/\tau}$ similar to the behavior of a capacitor in a DC circuit. However, if the circuit oscillates too fast, the capacitor remains somewhat charged between on and off cycles and its modulating effect on the circuit becomes smaller. For a combined DC and square-wave bias, the shift in the binding energy can be defined as:

$$V(t) = V_{DC} + V_{sq}\ e^{-t/\tau} \quad \text{Eq. 4}$$

where $\tau = R_{eq}\ C$ and $R_{eq} = R_1 + R_2 \sim 280$ MΩ. $V_{DC}$ is the magnitude of the DC voltage drop and was determined experimentally. $\tau$ and $R_{eq}$ are the time constant and the equivalent resistance of the circuit, respectively. Since XPS measures a time average of the signal, it is important to calculate the average voltage drop by integrating $V$(t) over half a period and normalizing by $T/2$.

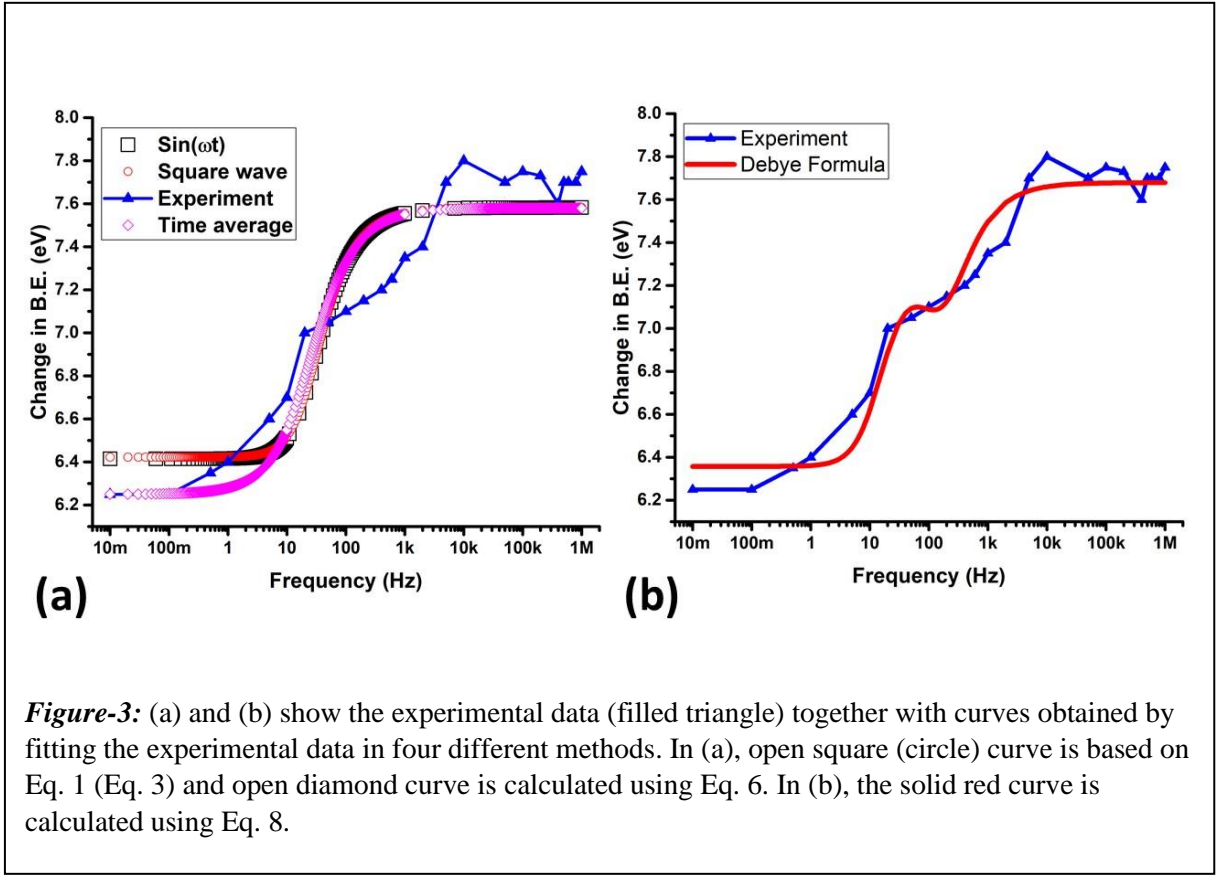

*Figure-3:* (a) and (b) show the experimental data (filled triangle) together with curves obtained by fitting the experimental data in four different methods. In (a), open square (circle) curve is based on Eq. 1 (Eq. 3) and open diamond curve is calculated using Eq. 6. In (b), the solid red curve is calculated using Eq. 8.

$$V_{avg} = \frac{2}{T}\int_0^{\frac{T}{2}}\left(V_{DC} + V_{sq}\,e^{-\frac{t}{\tau}}\right)dt = V_{DC} + \frac{2\tau A}{T}\left(1 - e^{-\frac{T}{2\tau}}\right) = V_{DC} + 2\tau V_{sq}f\left(1 - e^{-\frac{1}{2\tau f}}\right) \quad \text{Eq.5}$$

$$V_{avg} = V_{DC} + \tau V_{sq}\frac{\omega}{\pi}\left(1 - e^{-\frac{\pi}{\tau\omega}}\right) \quad \text{Eq.6}$$

We used Eq.6 to fit the experimental data with $V_{DC}$ fixed at $6.25\,V$ and $V_{sq}$ and $\tau$ extracted as $(1.33 \pm 0.048)$ V and $(0.0114 \pm 0.0033)s$. (Table-1, 3$^{rd}$ row) The time constant $\tau$ translates to a capacitance of $(41 \pm 12.5)$ pF when $R_{eq}$ equals $280\,M\Omega$. In Figure 3, the graph with open diamonds shows the fitted curve. Although this model is relatively simple, it accurately predicts the low-frequency behavior of the circuit and still manages to predict a sigmoidal curve with a similar slope predicted by previous methods. The fitted curve is also in line with the other two methods at the high-frequency region.

The mathematical models mentioned above cannot accurately reproduce the slope of the experimental curve from 1 Hz to 1 kHz. We attribute this to the fact that in all the models the capacitance of the thin film was assumed to be a constant. In reality, the dielectric constant and therefore the capacitance of a dielectric material exhibit frequency dependence. [21–24] Although there are more elaborate and accurate models to describe complex permittivity[21–23], we use the Debye formula to calculate the frequency dependence of the capacitance. The one exponential behavior in the time domain or the Debye formula, $\varepsilon(\omega) = \varepsilon_\infty + \frac{\Delta\varepsilon}{1+i\omega\tau}$, in the frequency domain gives an adequate description of the behavior of the orientation polarization for a large number of condensed systems. $\Delta\varepsilon = \varepsilon_s - \varepsilon_\infty$ is the dielectric strength related to the molecular dipole moment. $\varepsilon_s$ and $\varepsilon_\infty$ are the static and infinite-frequency permittivity, respectively. $\varepsilon_s$ for Si is 11.3.[25] Since the capacitor is directly proportional to permittivity which is the real part of Debye's formula $\text{Re}(\varepsilon(\omega)) = \varepsilon_\infty + \frac{\Delta\varepsilon}{1+(\omega\tau)^2}$

$$C = \text{Re}(\varepsilon(\omega))\frac{C_S}{\varepsilon_s} = \left(1 + \frac{10}{1+(\omega\tau)^2}\right)\frac{C_S}{\varepsilon_s} \qquad \text{Eq. 7}$$

where $C_S$ is the capacitance when a DC voltage is applied.

By combining Eq. 1 and Eq.7, we can derive an equation that correlates frequency dependence of the capacitance to the B.E. shift in XPS:

$$\Delta V_\omega = 10\text{ V} - I\left(R_1 + \frac{1}{\sqrt{\frac{1}{R_2^2}+(\omega C)^2}}\right) = 10\text{ V} - I\left(R_1 + \frac{1}{\sqrt{\frac{1}{R_2^2}+\left(\omega\left(1+\frac{\varepsilon_s-1}{1+(\omega\tau)^2}\right)\frac{C_S}{\varepsilon_s}\right)^2}}\right) \qquad \text{Eq. 8}$$

We used Eq. 8 to fit the experimental data. (Table-1, 4th row) In Figure 3b, the graph with a solid red line shows the fitted curve. The fitting looks better than the first three methods. The equivalent resistance, $R_{eq}$, extracted using Eq.8 is approximately equal to the experimentally determined value. However, the dipole relaxation time (τ) is significantly longer than typical oxides.[26] For close-packed quantum dot films, the high surface-to-volume ratio results in a higher relative number of polarized quantum dot-ligand bonds at the surface than in conventional systems which can lead to the dielectric constant of quantum dots exhibiting frequency dependency at much lower frequencies. [27,28] In addition to that, Debye formula is quite basic and may not be capable of addressing dynamics of dipole relaxation under X-ray and low-energy electron exposure.

## Conclusion

XPS measurements and an external circuit model have been successfully employed to obtain the electrical properties of SiNCs thin film. By constructing a DC external circuit with varying resistors during XPS measurements and monitoring the shift in binding energy for the Au 4*f* peaks, the equivalent resistance of the thin film of SiNCs was obtained experimentally. An application of an AC square pulse during XPS measurements exhibits the saturation of the frequency dependence of the shift in binding energy for the Si 2*p* peaks at very small and very big frequencies resulting in a sigmoidal response curve. We used three different models to extract the resistance and capacitance of the thin film. Table-1 summarizes all the parameters extracted using the models described above. An important point is that the capacitances agree well with each other within the standard error of each model. Sobolev et. al. measured the capacitance of the 1.2 nm Si nanoparticles on a Si substrate using a more traditional C-V method and found that the capacitance of thin film of SiNCs was ~20 pF when there is no forward or backward bias.[18] Equivalent resistance calculated using an AC circuit model matches almost perfectly with Sobolev's data. None of the models with constant capacitance accurately predict the slope of the experimental curve from 1 Hz to 1 kHz. In reality the dielectric constant and therefore the capacitance of a dielectric material exhibit frequency dependence. Therefore, we used Debye formula to add frequency dependence to the capacitance of the thin film. The fitting of the experimental data improved and the equivalent resistance extracted using Eq.8 becomes matches to the experimentally determined value indicating that the adding frequency dependence improved the model. However, the resistance and capacitance values extracted from XPS measurements are likely to be different than those measured via other methods such as scanning capacitance microscopy.

The main reason for any deviation can be attributed to electron dynamics under X-ray and low-energy electron exposure. A systematic and comparative study of the dielectric properties of materials can lead to a better understanding of the physics of these systems under X-ray and low-energy electron exposure.

## Acknowledgment

We thank the University of North Dakota for the support of this research.